\documentclass[preprint,proceedings]{rmaa}

\suppressfulladdresses 

\usepackage{paralist}

\usepackage{psfrag,color}

\newcommand{\msun}{\mbox{$M_{\odot}$}}
\newcommand{\be}{\begin{equation}}
\newcommand{\ee}{\end{equation}}
\newcommand{\noi}{\noindent}

\SetYear{2006}
\SetConfTitle{ Circumstellar Media and Late Stages of Massive Stellar Evolution}

\title{Supernova Explosions in Winds and Bubbles, with Applications to SN 1987A} 

\author{
  Vikram V. Dwarkadas,\altaffilmark{1} }

\altaffiltext{1}{Dept.~of Astronomy and astrophysics, Univ of Chicago,
5640 S Ellis Ave, AAC 010c, Chicago IL 60637 (vikram@oddjob.uchicago.edu).}

\shortauthor{Dwarkadas}
\shorttitle{SN, Winds and SN 1987A}

\listofauthors{V. V. Dwarkadas}
\indexauthor{Dwarkadas, V. V.}

\abstract{Massive stars can significantly modify the surrounding
medium during their lifetime. When the stars explode as supernovae,
the resulting shock wave expands within this modified medium and not
within the interstellar medium. We explore the evolution of the medium
around massive stars, and the expansion of the shock wave within this
medium. We then apply these results to understanding the expansion of
the shock wave in the ambient medium surrounding SN 1987A, and the
evolution of the radio and X-ray emission in this case.}

\addkeyword{hydrodynamics}
\addkeyword{shock waves}
\addkeyword{circumstellar matter}
\addkeyword{supernovae: general}
\addkeyword{supernovae: individual (SN 1987A)}

\begin{document}
\maketitle

\section{Introduction}
\label{sec:intro}

Core-collapse supernovae (SNe) arise from massive stars ($ \gtrsim 8
\msun$). These stars tend to modify the medium around them
substantially, often leading to the formation of circumstellar (CS)
wind-blown bubbles around the star. At the end of its life when the
star explodes as a SN, the resulting shock wave will interact with the
modified environment before it reaches the interstellar medium (ISM)
which in some cases may take a substantially long time. The subsequent
evolution of the shock wave, and the radiation signatures from the
supernova remnant (SNR) will depend crucially on the structure and
dynamics of the surrounding medium. In this paper we examine the
structure of the CS environment around stars, and the aftermath of
stellar explosions which occur within this environment. Finally we
apply the lessons learned to one of the most fascinating and
well-studied objects, SN 1987A. In \S \ref{sec:winds} we study the
formation of the ambient medium around massive stars. In \S
\ref{sec:SNe} we discuss the evolution of SNe in this medium. Finally
in \S \ref{sec:87a} we focus on applying these results to
understanding the evolution and radiation signatures from SN 1987A.

\section{The Environments of Massive stars}
\label{sec:winds}

The environment of a massive star depends on its zero-age
main-sequence mass, its rotational velocity, the metallicity, and the
presence of other nearby stars, among other factors. Therefore the
evolution can be quite complicated, but several common factors
exist. There has been significant discussion of the evolution of
massive stars by Norbert Langer at this conference, and a highly
comprehensive review of observations of the circumstellar medium (CSM)
around massive stars was given by You-Hua Chu. Therefore I will
concentrate here on those details that are essential to understanding
the evolution of the subsequent SN shock wave upon the explosion of the star.

\subsection{Main-Sequence Stars} Massive stars generally start their lives 
as main-sequence O/B stars. A star of solar metallicity during its
lifetime will usually lose mass through a radiatively driven wind
(Kudritzki \& Puls 2000) with a mass-loss rate on order 10$^{-8}$ to
10$^{-6} \msun$ yr$^{-1}$, and with a wind velocity on order a few
thousand km s$^{-1}$. The interaction of the wind from the star with
the surrounding medium will lead to the formation of an interstellar
wind-blown bubble (WBB), the structure of which was first delineated
by Weaver et al.~(1977, hereafter W77). If the wind parameters are
fixed, the bubble shows essentially 4 different regions as we go
outwards in radius (1) Freely expanding wind. For a given mass-loss
rate $\dot{M}$ and wind velocity v$_w$ the density of the wind goes as
$\rho_w = \dot{M}/(4 \pi r^2 v_w)$. (2) A low-density, high-pressure
(and therefore high temperature) region of shocked stellar wind. (3) A
region of shocked ambient medium. The shocked ambient medium usually
cools quickly, resulting in the formation of a thin, dense shell. (4)
The unshocked ambient medium. Most of the volume is usually occupied
by the region of shocked stellar wind, forming a hot, low-density
cavity. W77 found a self-similar solution for the bubble with constant
wind parameters, with radius varying with time as R $\propto t^{3/5}$.

Several factors may complicate this simplistic description. The wind
properties are not constant but change with time as the star evolves
through different phases. Hydrodynamic instabilities and the onset of
turbulence may cause a change in the dynamical properties. Mixing at
the interface between the hot shocked wind and the dense, cool,
shocked ambient medium, which may also be a conductive interface,
tends to lower the temperature in the interior. Density
inhomogeneities or an asymmetric wind may lead to bubbles that are not
spherical.

The radius of the outer shock of the bubble (the outer edge of the
dense shell) is (Weaver et al. 1977):
\be
\label{eq:rbub}
R_{sh} = 0.76 ~ \left( {L \over \rho} \right)^{1/5} t^{3/5}
\ee

\noi
where $L = 0.5 \dot{M} v_{w}^2 $ is the mechanical luminosity of the
wind, $\dot{M}$ is the wind mass-loss rate and $v_{w}$ is the wind
velocity. In the case of a main-sequence (MS) star with $\dot{M} =
10^{-7} \msun$ yr$^{-1}$, and velocity 2500 km s$^{-1}$, $L$ is about
1.984 $\times$ 10$^{35}$ ergs s$^{-1}$. If this lasts for about 10
million years, the total energy released will be about
\be E_{MS} = 6.25 \times 10^{49}\; {\dot{M}_{-7}}\; v^2_{2500} \; {t_{10}}
~~~ {\rm ergs}
\ee

\noindent
where ${\dot{M}_{-7}}$ is the mass-loss rate in terms of
10$^{-7}\msun$ yr$^{-1}$, ${v_{2500}}$ is the wind velocity in units
of 2500 km s$^{-1}$ and time is in units of 1 million years years
($t_{10}$ = 10 $\times 10^6$ years).

We assume that the main sequence star is formed in a medium with an
average density of about 2.34 $\times$ 10$^{-23}$ g cm$^{-3}$ (a
number density $\sim 10$ particles cm$^{-3}$, appropriate for an ionized
region). From equation \ref{eq:rbub} the radius of the swept-up shell
will be
\be
\label{eq:rbubms}
R_{MS} = 48.8 ~{\dot{M}_{-7}^{1/5}} \; v^{2/5}_{2500} \; {\rho}_{10}^{-1/5} \;
{t_{10}}^{3/5}~~ {\rm pc}
\ee

\noindent
where $\rho$ is in units of 2.34 $\times$ 10$^{-24}$ g cm$^{-3}$.
However we note that often, especially for less massive stars whose
lifetime is large, the bubble will come into pressure equilibrium with
its surroundings at an early stage, and thereafter will stall. The
radius of the bubble may then be up to an order of magnitude smaller
than is given in equation \ref{eq:rbubms} (see Dwarkadas 2006c, 2007).

The ratio  S$_m$ of the mass swept up by the shell to the total mass in the
wind is given by:
\be
S_m = \left( M_{sh} \over M_{wind} \right) = {4 \pi \over 3}
{R_{sh}^3 {\rho}_a \over \dot{M} t} = {4 \pi \over 3} ~ {0.76^3 \over 2^{3/5}} ~ \left( {{v_w}^3 t^2 {\rho} \over \dot{M}} \right)^{2/5}
\ee

\noindent
where ${\rho}_a$ is the ambient density. For the MS this becomes
\be
\label{eq:mswept}
S_m = 1.54 \times 10^5 ~    v^{6/5}_{2500} \; t_{10}^{4/5} ~ {\rho}_{10}^{2/5} \;
 {\dot{M}_{-7}^{-2/5}}
\ee

The wind velocity decreases slowly so the average velocity over the
entire main sequence phase could be lower than that used. The duration
of the main sequence phase also depends on the mass of the star,
decreasing with increasing mass, and the mass-loss rate may vary,
becoming increasingly higher as the MS phase comes to and end. Also,
as noted above, in several cases the bubble will not reach this radius
but may stagnate at a smaller radius. Due to all these reasons the
ratio given in \ref{eq:mswept} may be reduced by up to a couple of
orders of magnitude. However clearly the mass swept up by the shell
significantly dominates over the mass expelled in the wind, and the
wind material is not significant for the dynamics. Due to the large
mass of the swept-up shell compared to that of the wind, the
subsequent evolution is generally contained within the main-sequence
bubble. Thus, the radius of the bubble does not change significantly
after the main-sequence stage, although the internal structure may
undergo substantial change.

The swept-up mass lies in a thin, dense shell surrounding the bubble
cavity, which consists of an inner freely expanding wind region
followed by a region of almost constant density. The total wind mass
ejected over time $t$ is $\dot{M} t$. The mass in the freely expanding
wind is $\dot{M} R_t/v_w$ ($R_t$ = radius of wind termination
shock). Since $v_w ~t >> R_t$ by the end of the MS phase, a lower
limit to the average cavity density is obtained by assuming that the
wind material is uniformly distributed:
\be
\label{eq:bubden}
{\rho}_{bub} = {3 \dot{M} t \over 4 \pi R_{sh}^3} = {3 \over {4 \pi ~ 0.76^3}}
\left( 2 \dot{M}^{2/3} \rho_a \over v_w^2 \right)^{3/5} t^{-4/5}
\ee

\noi
which for the main sequence stage can be written as
\be
\label{eq:bubdennum}
{\rho}_{bubMS} = 1.5 \times 10^{-28}~ {\dot{M}_{-7}^{2/5}}\;
{\rho}_{10}^{3/5} \; v_{2500}^{-6/5}~ {t_{10}}^{-4/5}~~ {\rm g~ cm^{-1}}
\ee

Although this is a lower limit, especially if the bubble stalls early
in the MS phase, it can be seen that the density in the interior of
the bubble is on the order of $10^{-4}-10^{-3} $ particles cm$^{-3}$,
orders of magnitude lower than that of the surrounding medium.

The position of the wind-termination shock is of interest. An accurate
calculation is provided in Weaver et al.~1977 and Chevalier \& Imamura
1983. To obtain a simple approximate expression we assume that the
shock is strong, the shock jump is a factor of 4, and the bubble
density is approximately constant at the post-shock value. In order to
accommodate the results of W77, which indicate that the density in the
shocked wind increases by about a factor of 2 close to the contact
discontinuity, we include a factor $\alpha$, with $1 \la \alpha \la
2$. The total wind mass emitted over time $t$ is $\dot{M} t$, and the
mass of the freely expanding wind is $\dot{M} R_t / v_w$. The
difference between the two gives the total mass of the shocked wind
region between $R_c$ and $R_t$, where it is assumed that $R_c \approx
R_{sh}$:
\be
\label{eq:wtshk}
\dot{M} t - {{\dot{M} R_t} \over { v_w}} = {{4 \pi} \over {3}} (R_c^3 - R_t^3) {{4 \alpha \dot{M}} \over {4 \pi R_t^2 v_w}}
\ee

\noi
which gives
\be
{R_t \over R_c} = \left[ 1 + {3 \over {4 \alpha}}\left( {v_w \over R_t/t} - 1 \right) \right]^{-1/3}
\ee

Note that this equation is similar to equation 13 in Chevalier \&
Imamura 1983, if we take $M_2 >> 1$, and assume $s_2 = R_t/t$. For our
purposes this approximation is sufficient. We can simplify this
further by noting that in general $v_w t /R_t >> 1.$ Then we get:
\be
R_t = \left[ {4 \alpha R_c^3} \over {3 v_w t} \right]^{1/2}
\ee

For the parameters of the MS star, this gives
\be
R_t = 2.43 ~{\alpha}^{1/2} ~ v^{1/10}_{2500} \; {t_{10}}^{2/5} ~ {\rho}_{10}^{-3/10} \;
 {\dot{M}_{-7}^{3/10}} ~~ {\rm pc}
\ee

The small value of the wind termination shock shows that the hot
shocked wind occupies most of the volume of the main sequence bubble.

\subsection{Post Main Sequence Phases} 
Stars with M $ \leq 50 \msun$ generally become a red supergiant (RSG)
after leaving the MS. These stars have large envelopes and slow winds
(10-20 km s$^{-1}$) with a high mass loss rate of 10$^{-5}$ to
10$^{-4}~\msun$ yr$^{-1}$. This creates a high density region around
the star, confined by the pressure of the main-sequence bubble. For a
RSG lifetime of about 200,000 years the total energy input in the RSG
phase is
\be
E_{RSG} = 8 ~\times 10^{46} ~\dot{M}_{-4} ~v^2_{20} ~ t_{0.2} ~~ {\rm
ergs}
\ee

\noindent
where $\dot{M}_{-4}$ is the mass loss rate in terms of 10$^{-4}
{\msun}$ yr$^{-1}$, v$_{20}$ is velocity in units of 20 km s$^{-1}$,
and time is in units of million years, $t_{0.2} = 0.2 \times 10^6$
years.

Note that the RSG stage does not generally form a wind-blown bubble,
because the velocity of the RSG wind is much lower than that of the
medium (MS wind) into which it is blowing. However it does lead to a
new pressure equilibrium. Since the total ram pressure of the RSG wind
is different from that of the MS wind, the position of the wind
termination shock will change. The new position can be found by
equilibrating the ram pressure ($\rho v_w^2$) to the thermal pressure
in the MS bubble. According to W77 the thermal pressure in the MS
bubble is given by
\be
P_{bub} = 0.163 ~ L^{2/5} ~ {{\rho}_a}^{3/5} ~ t^{-4/5}
\ee

Equating to the ram pressure of the RSG wind gives the position of the
wind termination shock:
\begin{eqnarray}
{R_t}_{RSG} & = & \left[ {\dot{M_{RSG}} v_{RSG}} \over { 4 \pi ~ P_{bub}} \right]^{1
/2} \\
&  = & 8.85 ~ {\dot{M_R}_{-4}}^{0.5} ~ {{v_R}_{20}}^{0.5} ~ {\dot{M_{MS}}_{-7}}^{
-1/5} ~ \nonumber \\
& &{v_{MS}}_{2500}^{-2/5}~{{\rho_{10}}_a}^{-3/10}~{t_{MS}}_{10}^{2/5}~
{\rm pc}
\end{eqnarray}

\noi where in the last expression the subscript $R$ refers to the RSG
wind and the subscript $MS$ to the main sequence wind. It is not clear
however if the RSG wind will always be able to expand out to the
distance required to attain the new pressure equilibrium, and
therefore in some cases the pressure equilibrium may never be
attained.

The RSG wind with its low velocity expands a distance $R_{RSG}$, with
wind density $\rho_{RSG}$
\begin{eqnarray}
R_{RSG} & = & \kappa ~ 4.2 ~v_{20}~ t_{0.2}~ {\rm pc};\\
\rho_{RSG} & = & 2.81 \times 10^{-23}  ~\dot{M}_{-4}~v_{20}^{-1}~ r_{pc}^{-2}
\end{eqnarray}

\noi where we have added a factor $\kappa \ge 1 $ to account for the
fact that neither the transition, nor the change in velocity, from the
MS to the RSG phase is instantaneous. Given the size of the MS bubble,
it is clear that the RSG wind region will generally be confined to a
small fraction of the main-sequence bubble. The total mass lost during
the RSG phase is
\be
M_{RSG} = 20 ~ \dot{M}_{-4} ~ t_{0.2} ~~ \msun
\ee

Thus although total energy of the outflow in the RSG stage is small
compared to the MS stage and the subsequent Wolf-Rayet stage, a large
amount of stellar mass may be lost in the RSG stage.

Some stars may go from the RSG to a blue supergiant (BSG) stage, as in
the case of the progenitor of SN 1987A. The bipolar structure seen
around the object is often interpreted as resulting from the
interaction of a BSG and RSG wind. This is discussed further in \S
\ref{sec:87a}. The wind parameters in this case are uncertain, but the
mass-loss rate appears to be lower than even the MS stage, while the
wind velocity is intermediate between the MS and RSG stage.

Solar metallicity stars above 30-35 $\msun$ end their lives as
Wolf-Rayet (W-R) stars, although they may go through a luminous blue
variable (LBV) stage. The mass-loss decreases somewhat to about
$10^{-6}- 10^{-5}{\msun}$ yr$^{-1}$, while the wind velocity increases
to 2000 km s$^{-1}$. For a lifetime of 100,000 years, the total energy
input in the W-R phase is then \be E_{WR} = 4 \times 10^{49}~
\dot{M}_{-5}~ v^2_{2000} ~t_{0.1} ~{\rm ergs} \ee

Although the total mass is less, due to the high mass-loss rate and
wind velocity W-R winds may posses enough momentum to push out, and
possibly break up, any dense shell surrounding the star from the
previous intermediate wind stage, distributing its contents throughout
the surrounding medium. Generally they will have enough momentum to
collide with the MS shell, sending a reflected shock back. A W-R wind
termination shock will be formed where the thermal pressure of the
shocked wind bubble equals the ram pressure of the freely flowing
wind.

The post-MS stages may add considerable mass to the bubble without
increasing the volume much. However equations (\ref{eq:bubden}) and
(\ref{eq:bubdennum}) imply that, even increasing the mass ($\dot{M}
t$) by a factor of 30-40 results in number densities of order 10$^{-3}
- 10^{-2}$ cm$^{-3}$. Therefore the density over the bubble interior
is in general low for W-R bubbles.

Multidimensional calculation (Garcia-Segura et al.~1996; Freyer et
al.~2006; Dwarkadas 2006b) reveal the presence of hydrodynamic
instabilities in many stages. In one calculation of the medium around
a 35 $\msun$ star, Dwarkadas (2006b) found that both the RSG and
subsequent W-R wind (expanding into the RSG wind) were Rayleigh-Taylor
unstable (Fig 1a). These instabilities tend to break-up the RSG shell
and distribute its material over the entire bubble. They may lead to
the formation of blobs, clumps and filaments in the WBB. Due to
density fluctuations in various stages, the bubble interior becomes
turbulent by the end of the simulation.

Other factors will considerably modify this simple picture. Rotation
of the star can lead to an increase in the mass-loss rate (Maeder \&
Meynet 2000). Dwarkadas \& Owocki (2002) showed that a star rotating
close to its break-up velocity may emit a wind preferentially in the
polar direction. Mass-loss rates, which depend roughly as the square
root of the metallicity, may be reduced at lower metallicities,
although rotation can still lead to high rates. And the presence of a
binary companion can significantly alter the evolution, as has been
hypothesized for SN 1987A (Morris \& Podsiadlowski 2005).

\section{Supernova Evolution}
\label{sec:SNe}

The above description clearly illustrates a few salient points:

\begin{itemize}
\item The medium around most massive stars consists of a low density
cavity created in the main-sequence stage, surrounded by a dense
shell.
\item If the progenitor is a RSG then the density near the star may be
quite high, as expected for a RSG wind density. If the RSG stage is
followed by a Wolf-Rayet stage then this can distribute the RSG
material over the entire WBB, and the density will be much lower.
\item In either case the region close in to the star is usually a
freely flowing wind, with a wind density that generally decreases as
r$^{-2}$. However if the wind parameters change substantially with
time then the density dependence may change. Several authors have
suggested that the density of the medium surrounding SN 1993J
decreases only as r$^{-1.5}$ (Suzuki \& Nomoto 1995; Mioduszewski et
al.~2001; Bartel et al.~2002).
\end{itemize}

When the star explodes as a SN the resulting shock wave will first
expand in the surrounding wind medium, and then in the low density
cavity. If the surrounding wind medium is a RSG then the shock wave
will expand in a higher density medium, and therefore its luminosity,
due to circumstellar interaction (Chevalier \& Fransson 1994) will be
high. If the shock wave results from the death of a W-R star then the
surrounding medium density will be much lower, and the resulting
luminosity will also be lower.

The shock wave will expand in the wind until it reaches the wind
termination shock, after which it will continue to expand in the low
density shocked-wind medium. If it was earlier expanding in a RSG wind
then its luminosity will be expected to drop. If it was expanding in a
Wolf-Rayet wind then the luminosity will not change much, or may in
fact increase slightly as the shock wave crosses the wind termination
shock.

Since most of the mass is in the dense shell, the interaction of the
shock wave with the dense shell controls the relevant dynamics, which
depends mainly on one parameter $\Lambda$, the ratio of the mass of
the shell to that of the ejected material. Simulations for various
values of $\Lambda$ have been carried out by Tenorio-Tagle et
al.~(1990) and Dwarkadas (2005). They show that values of $\Lambda \ga
1$ are dynamically important. As the value of $\Lambda$ increases, the
kinetic energy transmitted from the ejecta to the shell gradually
increases. The collision with the shell results in a transmitted shock
expanding into the shell, and a reflected shock moving back into the
ejecta. It also results in an increase in the emission from the
remnant, especially the optical, X-ray and radio emission. The
reflected shock moves back towards the origin, thermalizing the ejecta
on its way, faster than the original SN reverse shock would have. For
smaller $\Lambda$ the SN shock wave eventually `forgets' about the
existence of the shell, and the solution resembles what it would be in
the absence of the shell. The density structure changes to reflect
this. An increase in $\Lambda$ results in higher velocity reflected
shocks, while the transmitted shock slows down considerably. In
extreme cases the transmitted shock may be trapped in the dense shell
for a significant amount of time, and the shock wave may lose
considerable energy to become a radiative shock wave. In such a case
it can go from the free-expansion stage to the radiative stage without
ever going through the Sedov or adiabatic phase. The remnant is also
confined to the shell, whose size as we have seen was set in the MS
stage.

The evolution of SNe in wind blown bubbles thus differs considerably
from the self-similar solutions so often used to describe SN evolution
in general. The radius and velocity of the remnant do not evolve in a
self-similar fashion but may vary considerably once the shock expands
beyond the freely-flowing wind. Furthermore the expansion parameter of
the remnant $\delta$ (where R $\propto$ t$^{\delta}$) is continuously
varying, as opposed to a self-similar case where the evolution is
constant. This is illustrated in Dwarkadas (2005).

In multi-dimensions (Dwarkadas 2006a,b) the interior of the nebula
shows signs of turbulence, with significant density and pressure
variations. The shock wave expanding in this medium no longer remains
spherical, but becomes wrinkled due to the continuous interaction with
the pressure and density inhomogeneities. It develops a corrugated
structure (see Fig 1b), and its interaction with the surrounding shell
no longer occurs all at once but in a piecemeal fashion, with some
parts of the shock colliding with the shell before others. As
mentioned before, the collision results in an increase in the
luminosity. In this case the luminosity of some parts of the shell
will increase first, followed by those in other parts of the shell. A
similar scenario appears to be taking place in the ring around SN
1987A. 

This brief description encapsulates the basic properties of the
evolution of SNe in the medium around massive stars. Further details
are given in Dwarkadas (2005, 2006a,b). In the rest of this paper we
would like to concentrate on applying the results described above to
SN 1987A.

\begin{figure}
\includegraphics[height=.18\textheight]{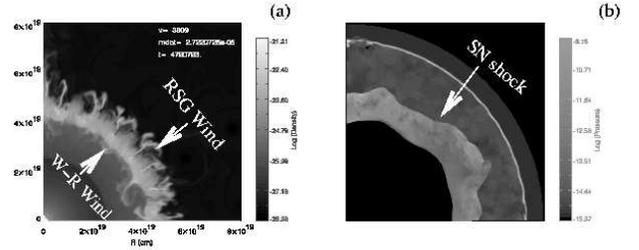}
\caption{(a) The evolution of the W-R wind into the RSG wind region
  for a 35 $\msun$ star. Both winds are R-T unstable. (b) The SN shock
  wave expanding into the density inhomogeneities in the interior can
  becomes wrinkled and lose its spherical shape (Dwarkadas 2006a,b).}
\end{figure}

\section{SN 1987A}
\label{sec:87a}

SN 1987A provides one of the best, and most spectacular, opportunities
to witness the evolution of a SN shock wave within a wind-blown
bubble. The three ring structure around SN 1987A has been interpreted
as an hour-glass shape bubble formed by the interaction of a
blue-supergiant (BSG) and a RSG wind (Blondin \& Lundqvist 1993;
Sugerman et al.~2005). The relatively small size of the equatorial
ring (possibly the waist of a wind-blown nebula) has ensured that we
can see the shock-shell collision unfold in a short time span after
the SN explosion. SN 1987A has allowed us to refine our theories of
SNe evolving in the winds of massive stars, while providing
confirmation of many existing ideas, and catalyzing many new ones.

Chevalier \& Dwarkadas (1995) showed that the slow expansion of the
radio source, its large size, and the almost linear increase in radio
and X-ray emission could be understood if we assumed that the SN shock
wave was interacting with an HII region inside of the inner ring. In
the years since this model has been verified by Dick McCray and his
collaborators (Michael et al.~1998). This model modifies the picture
outlined above by including a high-density HII region inside of the
outer shell, in the low density wind-driven cavity. A depiction of the
surrounding medium into which the SN explodes (in the equatorial
region) can be seen in Fig.~2.

The morphology and dynamics of the ejecta in SN 1987A are inherently
three-dimensional. The gradual appearance of bright spots around the
equatorial ring (Lawrence et al.~2000), and the changes in the X-ray
and optical luminosity indicate an aspherical shock interacting with
several protrusions, formed perhaps by instabilities, emanating from
the equatorial ring. All this presents a very complex morphology for
the modeler.

Yet there are several aspects of SN 1987A that can be understood by
resorting to a simple, spherically symmetric model. This includes the
radius and velocity of the expanding shock wave, the re-emergence of
the X-ray and radio emission at around 1100 days after explosion, and
the gradual increase in the X-ray emission. We present here a toy
model, a spherically symmetric calculation that captures the basic
idea of Chevalier \& Dwarkadas (1995), and compute from this the shock
dynamics, the hard X-ray emission and radio emission, which we compare
with observations. The hydrodynamic simulations have been carried out
using the VH-1 code, a spherically symmetric multi-dimensional
finite-difference hydrodynamic code.

\begin{figure}[t]
\includegraphics[height=.25\textheight]{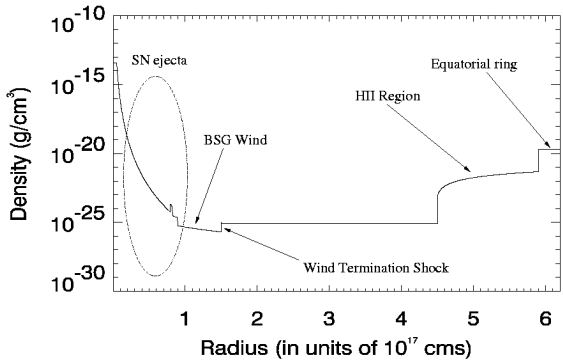}
\caption{The density profile just after the start of the
simulation. The various regions (free expanding wind, almost constant
density bubble, HII region and equatorial ring) are marked.  }
\end{figure}

Figure 2 shows the initial conditions for the run. The SN ejecta, the
wind termination shock of the BSG wind, the HII region and the
equatorial ring are shown. The various parameters used for the run
were: BSG wind mass-loss rate $\dot{M} = 5 \times 10^{-9} \msun$/yr,
BSG wind velocity = 550 km s$^{-1}$, radius of wind termination shock
= 1.5 $\times 10^{17}$ cm, inner radius of HII region = 4.5 $\times
10^{17}$ cm, inner radius of equatorial ring = 5.9 $\times 10^{17}$
cm. The wind density is obtained from $\rho_w = \dot{M}/(4 \pi r^2
v_w)$. The density is assumed to jump a factor of 4 at the wind
termination shock, after which it remains constant up to the HII
region, where the density jump is a factor of 100. The HII region
density is assumed to increase steadily until the inner ring radius,
whose density starts at 10$^{-20}$ g cm$^{-3}$ and increases
linearly. The SN ejecta density varies as r$^{-9}$. These parameters
were initially selected based on observational and theoretical work by
Lundqvist (1998), Michael et al.(1998), Manchester et al.~(2002), Park
et al.~(2006), and then modified to obtain a better fit to the
observations.

\begin{figure}[t]
\includegraphics[height=.6\textheight]{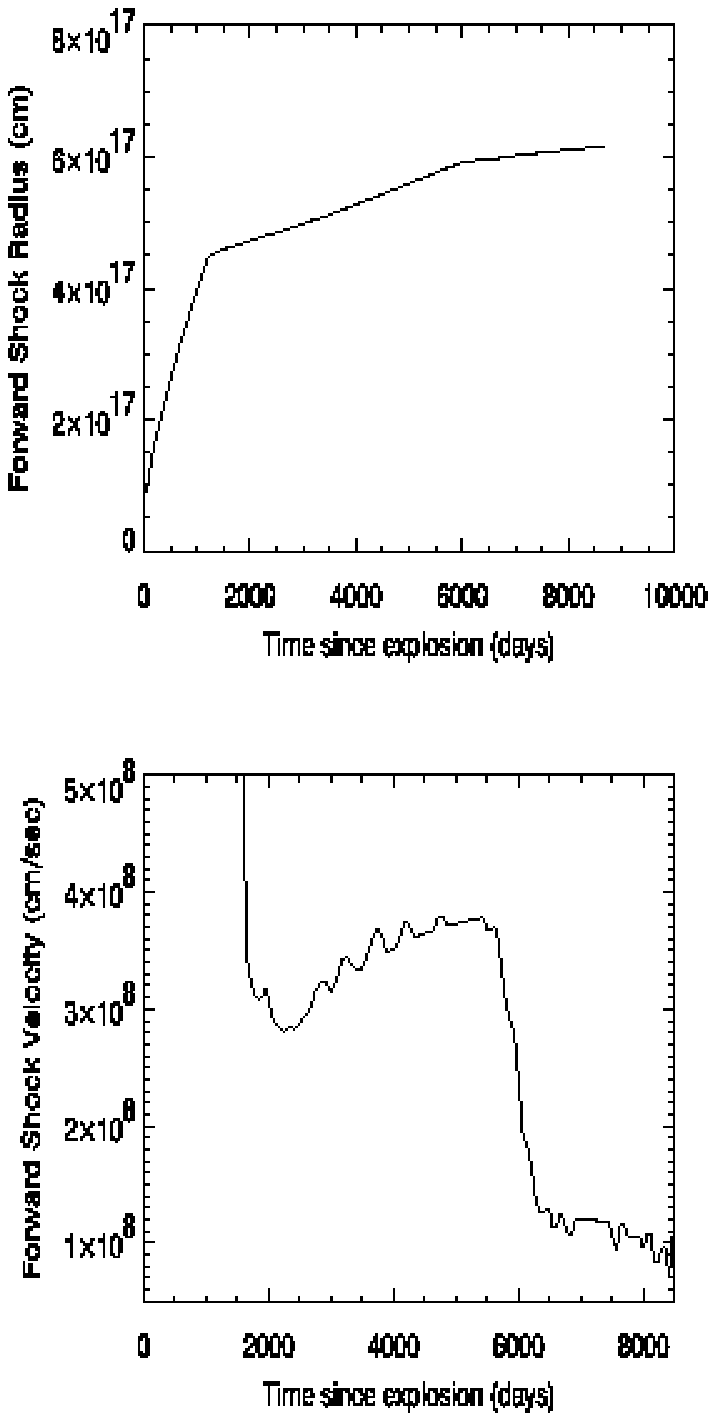}
\caption{The time evolution of the radius (top) and velocity (bottom)
of the SN forward shock wave. }
\end{figure}

In a future paper we will discuss the detailed hydrodynamics of the
evolution. Herein we concentrate on the main results. The radius and
velocity of the forward shock wave are shown in Fig 3. These
quantities are in good agreement with 2 important observations: (1)
Radio data from Manchester et al.~(2002) which suggests that the shock
velocity reduces abruptly from a large value ($> 35000$ km s$^{-1}$)
to a value of around 3000 km s$^{-1}$ around day 1100; and (2) X-ray
observations (Park et al.~2005, 2006) which indicate that around day
6000-6200 the shock velocity decreased to about 1500-1600 km s$^{-1}$.

Figure 4 shows an approximate computation of the radio emission from
the remnant. We follow the prescription of Chevalier (1982), assuming
that the radio emitting region lies between the forward and reverse
shocks, with the optically thin radio luminosity being given by 
\be
\label{eq:radiolum}
L_{\nu} \propto 4 \pi R^2 {\Delta} R K B^{(\gamma + 1)/2}
{\nu}^{(\gamma - 1)/2} 
\ee

\noindent
where $L_{\nu}$ is the radio luminosity at frequency $\nu$, ${\Delta}
R$ is the thickness of the synchrotron emitting region, $B$ is the
magnetic filed, and the distribution of accelerated particles is
assumed to be a power-law $N(E) = K E^{-{\gamma}}$. We further assume
that both $B$ and $K$ scale simply as the hydrodynamic variables: 
\be
\label{eq:scaling}
U_B = {\epsilon}_{B} U_{th}; ~~~~~~~~~~~~ U_{rel} = {\epsilon}_{r}
U_{th} 
\ee

\noindent
where $U_B$ is the magnetic energy density, $U_{rel}$ is the
relativistic particle energy density, and $U_{th}$ is the thermal
energy density.  The value of $\gamma$ is taken to be 2.7 (Manchester
et al.~2002).

\begin{figure}
\includegraphics[height=.27\textheight]{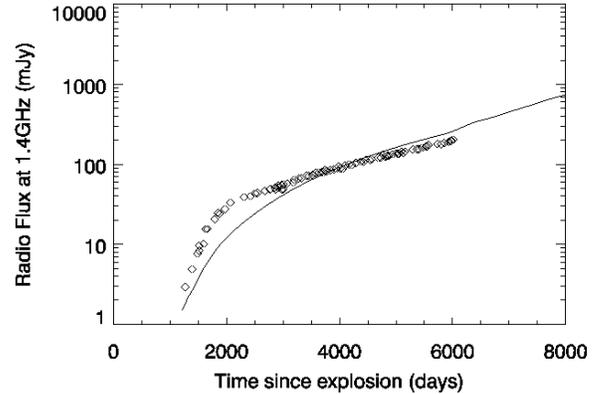}
\caption{The evolution of the radio luminosity with time, assuming that
the parameter $K$ is constant.}
\end{figure}

In Fig 4 the solid line shows our prediction of the radio luminosity,
while the symbols represent the observed data at 1.4 GHz (Manchester
et al.~2002). While the fit over the observed range is within a factor
of a few, we find it difficult to get the correct linear slope. In our
calculations the emission is usually increasing quadratically or
faster, and by the time the shock hits the ring the radio emission
begins to increase at an increasingly fast pace, which is not found in
the observations. In our exploration of the parameter space we have
not found a suitable set of parameters which generate a linear
increase in radio luminosity.

This model is unusual in that the best fit is obtained when either the
magnetic field or the parameter $K$ representing relative particle
energy density is assumed constant, with a slight preference for the
latter. The reason why either quantity being constant works is due to
the scaling in equation \ref{eq:radiolum}. The important parameter is
the quantity $K B^{(\gamma + 1)/2}$. With $\gamma \sim 3$, this
parameter becomes $K B^2 $. Coupled with the scaling in equation
\ref{eq:scaling} this implies that either $K$ constant and $B \propto
P^{0.5}$, or $B$ constant and $K \propto P$, both give $K B^2 \approx
U_{th} \propto P $, where $P$ is the thermal pressure. Thus this
behavior is simply a result of $\gamma$ being close to 3. It is not
clear why a constant value for either quantity works best.

Figure 5 shows a comparison of the hard X-ray luminosity computed
using the CHIANTI code (Dere et al.~1997, Landi et
al.~2006). According to Park et al.~(2005,2006) the soft xray emission
may be coming from the interaction of the shock wave with the
protrusions in the ring. Furthermore line emission forms a strong
component of the soft xrays. These computations are beyond the scope
of this paper, and we have concentrated only on the hard xray
emission. We also assume that, for the given shock speed, the electron
temperature is much lower than the ion temperature by a factor of 50
(see Ghavamian et al.~2006). This provides a reasonable fit to the
X-ray emission. The solid line in Fig.~5 is the emission calculated
from our simulations, the data points are from Park et al.~2006. The
comparison is reasonably good before the shock hits the ring, but
there is a significant bump in the luminosity upon shock-ring
interaction, something that is not seen in the data. Most of the hard
X-ray emission appears to arise from the reverse-shocked ejecta.

\begin{figure}
\includegraphics[height=.27\textheight]{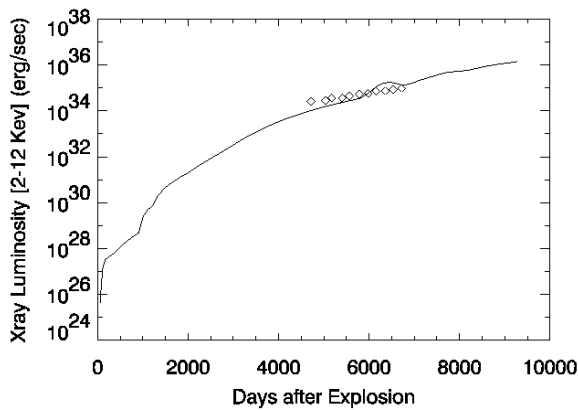}
\caption{The variation of the hard X-rays with time. T$_e$ = 0.02
T$_i$, where 'e' = electrons and 'i' = ions. The emission is mostly
coming from the reverse-shock region. }
\end{figure}

There are several shortcomings to this simple spherically symmetric
approach, and the subsequent computation of the radio and X-ray
emission. Relativistic electrons may not be injected in a power-law
fashion, the magnetic and/or relativistic particle energy density may
not vary as the hydrodynamic variables, or the variations may be
different as the shock traverses the various density structures. The
spectral index varies somewhat over the evolution (Manchester et
al.~2002). The expansion is not spherically symmetric, so a
spherically symmetric model may not necessarily work well. There is
the possibility that with aspherical expansion light travel time
effects may come into play. The ratio of the electron to ion
(post-shock) temperature, assumed constant here, may vary throughout
the evolution. There will be contribution to the Xrays from the shock
wave hitting the protrusions from the equatorial ring, which will be
discussed in a more comprehensive model in future. We cannot hope to
get much better agreement from such a simplified model, and it is
satisfying that it does give reasonably good agreement with the
observed data.

In a future paper, we will discuss in more detail the hydrodynamics of
the evolution and the reverse shock dynamics, which space constraints
do not permit us to include herein. We will investigate more
thoroughly the emission from different regions, as well as the soft
xray emission. What is also needed is a direct computation of the
medium around SN 1987A, including the ionization from the star and the
formation of the HII region, at least in the equatorial plane. These
will be addressed in future.

\begin{acknowledgments}
VVD is supported by award \# AST-0319261 from the National Science
Foundation, and by NASA through grant \# HST-AR-10649 awarded by the
Space Science Telescope Institute. I would like to thank the
organizers, and Guillermo Garcia-Segura in particular, for an
extremely broad and interesting conference, and for their gracious
hospitality. Very helpful conversations with Roger Chevalier and
Sangwook Park are greatly acknowledged.
\end{acknowledgments}

\end{document}